\documentclass[useAMS,usenatbib]{mn2e}
\usepackage{epsfig}
\usepackage{float}
\usepackage[T1]{fontenc}
\usepackage{graphicx}
\usepackage{epstopdf}

\usepackage{natbib}

\usepackage{color}

\usepackage{aas_macros}

\usepackage{array}
\newcolumntype{L}[1]{>{\raggedright\let\newline\\
\arraybackslash\hspace{0pt}}m{#1}}
\newcolumntype{C}[1]{>{\centering\let\newline\\
\arraybackslash\hspace{0pt}}m{#1}}
\newcolumntype{R}[1]{>{\raggedleft\let\newline\\
\arraybackslash\hspace{0pt}}m{#1}}

\def\Re{\mbox{$R_{\rm e}$}}

\def\Msun{\mbox{$M_\odot$}}

\def\Mdyn{\mbox{$M_{\rm dyn}$}}

\def\mst{\mbox{$M_{\star}$}}

\def\lsim{\mathrel{\rlap{\lower3.5pt\hbox{\hskip0.5pt$\sim$}}
    \raise0.5pt\hbox{$<$}}}                % less than or approx. symbol
\def\gsim{~\rlap{$>$}{\lower 1.0ex\hbox{$\sim$}}}

\def\SN{\mbox{$S/N$}}
\def\zspec{\mbox{$z_{\rm spec}$}}
\def\zphot{\mbox{$z_{\rm phot}$}}
\def\Mauto{\mbox{{\tt MAG\_AUTO}}}
\def\Mautor{\mbox{{\tt MAG\_AUTO\_r}}}
\def\MErrautor{\mbox{{\tt MAGERR\_AUTO\_r}}}

\def\zs{\mbox{$z_{\rm spec}$}}

\def\Fig{\mbox{Fig.~}}
\def\Figs{\mbox{Figs.~}}

\def\Sec{\mbox{Sec.~}}

\def\magaptwo{\mbox{{\tt MAGAP\_2}}}
\def\magapfour{\mbox{{\tt MAGAP\_4}}}
\def\magapsix{\mbox{{\tt MAGAP\_6}}}
\def\amthree{\mbox{{\tt APERMAG\_3}}}

\def\gzero{\mbox{grade-0}}
\def\gone{\mbox{grade-1}}
\def\gtwo{\mbox{grade-2}}

\def\ggr{\mbox{$\nabla_{\rm g-r}$}}
\def\gri{\mbox{$\nabla_{\rm r-i}$}}
\def\ggi{\mbox{$\nabla_{\rm g-i}$}}

\def\CS{\mbox{{\tt C-Sample}}}
\def\CSs{\mbox{{\tt C-Samples}}}
\def\FULL{\mbox{{\tt FULL}}}
\def\ETGs{\mbox{{\tt ETGs}}}
\def\MSCGs{\mbox{{\tt MSCGs}}}
\def\MSCG{\mbox{{\tt MSCG}}}

\def\u{\mbox{$u$}}
\def\rband{\mbox{$r$}}
\def\g{\mbox{$g$}}
\def\i{\mbox{$i$}}
\def\J{\mbox{$J$}}
\def\Ks{\mbox{$Ks$}}

\def\Remaj{\mbox{$R_{\rm e, maj}$}}

\title[Super-compact galaxies]{Towards a census of super-compact massive galaxies in the Kilo Degree Survey}

\author[Tortora C. et al.]{\noindent
C.~Tortora$^{1}$\thanks{E-mail: ctortora@na.astro.it}, F.~La
Barbera$^{1}$, N.R.~Napolitano$^{1}$, N.~Roy$^{1,2}$,
M.~Radovich$^{3}$, \and S.~Cavuoti$^{1}$, M.~Brescia$^{1}$,
G.~Longo$^{2}$, F.~Getman$^{1}$, M.~Capaccioli$^{2}$,
A.~Grado$^{1}$, \and K.~H.~Kuijken$^{4}$, J.~T.~A.~de Jong$^{4}$,
J.~P.~McFarland$^{5}$, E.~Puddu$^{1}$
\\~\\
$^1$ INAF -- Osservatorio Astronomico di
Capodimonte, Salita Moiariello, 16, 80131 - Napoli, Italy\\
$^2$ Dipartimento di Scienze Fisiche, Universit\`{a} di Napoli
Federico II, Compl. Univ. Monte S. Angelo, 80126 - Napoli, Italy\\
$^3$ INAF -- Osservatorio Astronomico di Padova, Via Ekar, 36012
Asiago VI 0424 462032\\
$^4$ Leiden Observatory, Leiden University, P.O. Box 9513, 2300 RA
Leiden, the Netherlands\\
$^5$ Kapteyn Astronomical Institute, University of Groningen, P.O.
Box 800, 9700 AV Groningen, the Netherlands}

\begin{document}
\date{Accepted  Received }
\pagerange{\pageref{firstpage}--\pageref{lastpage}} \pubyear{xxxx}
\maketitle

\label{firstpage}
\begin{abstract}
The abundance of compact, massive, early-type galaxies (ETGs)
provides important constraints to galaxy formation scenarios.
Thanks to the area covered, depth, excellent spatial resolution
and seeing, the ESO Public optical Kilo Degree Survey (KiDS),
carried out with the VLT Survey Telescope (VST), offers a unique
opportunity to conduct a complete census of the most compact
galaxies in the Universe. This paper presents a first census of
such systems from the first 156 square degrees of KiDS. Our
analysis relies on g-, r-, and i-band effective radii (\Re ),
derived by fitting galaxy images with PSF-convolved S\'ersic
models, high-quality photometric redshifts, \zphot, estimated from
machine learning techniques, and stellar masses, \mst, calculated
from KiDS aperture photometry. After massiveness ($\mst \gsim 8
\times 10^{10}\, \rm \Msun$) and compactness ($\Re \lsim 1.5 \,
\rm kpc$ in \g-, \rband- and \i-bands) criteria are applied, a
visual inspection of the candidates plus near-infrared photometry
from VIKING-DR1 are used to refine our sample.  The final catalog,
to be spectroscopically confirmed, consists of 92 systems in the
redshift range $z \sim 0.2-0.7$.  This sample, which we expect to
increase by a factor of ten over the total survey area, represents
the first attempt to select massive super-compact ETGs (\MSCGs) in
KiDS.  We investigate the impact of redshift systematics in the
selection, finding that this seems to be a major source of
contamination in our sample.  A preliminary analysis shows that
\MSCGs\ exhibit negative internal colour gradients, consistent
with a passive evolution of these systems.  We find that the
number density of \MSCGs\ is only mildly consistent with
predictions from simulations at $z>0.2$, while no such system is
found at $z < 0.2$.
\end{abstract}

\begin{keywords}
galaxies: evolution  -- galaxies: general -- galaxies: elliptical
and lenticular, cD -- galaxies: structure.
\end{keywords}

\section{Introduction}\label{sec:intro}

The understanding  of the  physical processes  which drive  the
galaxy mass built up  and size accretion are among the  most
timely topics in galaxy evolution studies. Massive early-type
galaxies (ETGs) are found to be much more compact in the past than
in the present Universe (\citealt{Daddi+05};
\citealt{Trujillo+06};   \citealt{Trujillo+07};
\citealt{vanderWel+08}).   At  redshifts   $z>2$,  while  the
massive star-forming   disks    have   effective   radii   of
several   kpc (\citealt{Genzel+08}), the  quenched spheroids
(``red  nuggets'') have small effective  radii of about 1  kpc.
Such red nuggets  are thought to  form   through  a   chain of
different processes:   a) accretion-driven violent disc
instability, b) dissipative contraction resulting in the
formation  of compact,  star-forming ``blue nuggets'', c)
quenching of star formation (\citealt{Dekel_Burkert14}). After
these  processes occur, a gradual  expansion in size of  the red
nuggets may take  place, leading to the formation of  the massive
ETGs we  observe in  the  nearby Universe.   Theoretical  studies
point  to dry-mergers as  the dominant mechanism  for the size and
stellar mass growth        of       very        dense massive
galaxies (\citealt{Khochfar_Silk06}). In particular, minor mergers
would provide a modest  stellar mass accretion, but a strong
evolution in galaxy     size (\citealt{vanDokkum+10};
\citealt{Hilz+13}; \citealt{BNE14}; \citealt{Tortora+14_DMevol}).
Mergers are believed to be common for very massive systems at high
redshifts, with major merger rates (mergers per galaxy per Gyr) in
the range $0.3-1 \, \rm Gyr^{-1}$ at $z \sim 2$ and smaller than
$0.2 \, \rm Gyr^{-1}$ at $z  \lsim  0.5$
(\citealt{Hopkins+10_Mergers_LCDM}).   An alternative scenario
explains the  size evolution as the result of (e.g.)  quasar
feedback, rather than merging, making galaxies to puff up after a
lost of large amounts of (cold) gas (\citealt{Fan+08,Fan+10}).

Over cosmic time, one may expect that high-z compact galaxies
evolve into present-day, massive, big galaxies.   However, a
fraction of these objects might survive intact till the present
epoch, resulting in compact, relic systems in the nearby Universe
characterized by old stellar populations. Recently there have been
some efforts to search for massive compact galaxies at low
redshifts (\citealt{Trujillo+09_superdense};
\citealt{Taylor+10_compacts}; \citealt{Valentinuzzi+10_WINGS};
\citealt{Shih_Stockton11}; \citealt{Trujillo+12_compacts};
\citealt{Poggianti+13_low_z}; \citealt{Poggianti+13_evol};
\citealt{Damjanov+13_compacts, Damjanov+14_compacts,
  Damjanov+15_compacts}; \citealt{Saulder+15_compacts}) and
investigate further their dynamical and stellar population
properties, as well as the role of environment on their properties
(e.g. \citealt{Ferre-Mateu+12}; \citealt{Lasker+13_IMF_compact};
\citealt{Ferre-Mateu+15}; \citealt{Trujillo+14};
\citealt{Yildirim+15}; \citealt{Damjanov+15_env_compacts};
\citealt{Stringer+15_compacts}; \citealt{Wellons+15_lower_z}).

In some theoretical models, that include the effect of galaxy
mergers, the fraction of massive objects that would survive
without undergoing any significant transformation since $z \sim 2$
to the present could reach a fraction of about $1-10\%$
(\citealt{Hopkins+09_DELGN_IV}; \citealt{Quilis_Trujillo13}).  At
``low'' redshifts ($z \lsim 0.2$), theoretical models predict a
density of relic remnants in the range $10^{-7}-10^{-5}$, which
means that, in large surveys like the Sloan Digital Sky Survey
(SDSS), we might expect to find a few candidates in this redshift
range. However, recent observational works have shown the paucity
of old, super-compact ($\Re \lsim 1.5$ kpc), massive ($\mst \gsim
10^{11}\, \rm \Msun$) galaxies in the local Universe ($z \lsim
0.2$; \citealt{Trujillo+09_superdense};
\citealt{Taylor+10_compacts}; \citealt{Ferre-Mateu+12}).  Indeed,
NGC1277, a nearby lenticular galaxy in the Perseus cluster, is
actually the only well-characterized, old system at $z \sim 0$,
that might be a true relic galaxy (\citealt{Trujillo+14}; see also
\citealt{Martin-Navarro+15_IMF_relic}).  Other candidates have
been recently detected by \cite{Saulder+15_compacts}, although
only a few of them fulfill the above (restrictive) size and mass
criteria ($\Re \lsim 1.5$ kpc and $\mst \gsim 10^{11}\, \rm
\Msun$), none of them being at $z < 0.05$.  Larger numbers of old
compact systems have been found at lower masses ($< 10^{11}\, \rm
\Msun$), when relaxing the compactness selection criteria
(\citealt{Valentinuzzi+10_WINGS}; \citealt{Poggianti+13_low_z}).
The (almost) lack of nearby super-compact, relic systems may
represent a challenge for the current paradigm of galaxy
formation; in particular, we have to understand whether this lack
is due to some observational bias, e.g.  the limited spatial
resolution of photometric data at $z \sim 0$; a failure of
theoretical predictions; and/or an environmental effect, for
which, as suggested by NGC\,1277, relic galaxies might be more
frequent in high-density cluster regions.

In the intermediate redshift range ($0.2 \lsim z \lsim 0.7$),
compacts have been recently investigated in detail by
\cite{Damjanov+14_compacts}, who selected  $\sim 200$ massive
compacts from a sample  of stellar-like objects within the
6373.2~sq.~deg. of the BOSS survey; $20\%$ of these galaxies are
dominated by old stellar populations, which make them reliable
candidates to be the product of the unperturbed evolution of
compact high--z systems.  However, $93\%$ of  these  galaxies do
not  have  measured  \Re\, which  hampers  the selection  of  such
systems  as truly  compact  objects. More recently,
\cite{Damjanov+15_compacts},  have   analyzed  F814W  HST images
for the COSMOS field,  providing robust size measurements for a
sample of 1599 compact systems in the  range $0.2 \lsim  z \lsim
0.8$. Other studies have  performed  detailed analysis of stellar
populations and morphology of small samples of  compact galaxies
at these redshifts (\citealt{Hsu+14_compacts};
\citealt{Stockton+14_compacts}).  The population of dense
passively evolving  galaxies  in  this   intermediate  redshift
range  possibly represents  a link  between  compact systems,
dominating the  massive quiescent galaxy population at high  z,
and their descendants in the  nearby  Universe.   Indeed, large
samples  of  compacts,  with high-quality   photometry (to derive
reliable   structural   parameters),  and  spectroscopic data, are
actually  necessary  to better understand the formation and
evolution of these systems.

The Kilo Degree Survey (KiDS; \citealt{deJong+15_KiDS_paperI}) is
one of the ESO public surveys carried out with the VLT Survey
Telescope (VST; \citealt{Capaccioli_Schipani11}) equipped with the
one square degree field of view and high angular resolution
($0.2''/pixel$) OmegaCAM camera (\citealt{Kuijken+04};
\citealt{Kuijken11}). KiDS is mainly designed for weak lensing
studies, providing deep imaging in four optical bands ($ugri$),
over a 1500 square degree of the sky with excellent seeing (e.g.
$0.65''$ median FWHM in \rband-band). The high image quality and
deep photometry are ideal to investigate massive compact systems.

According to predictions from simulations (\citealt{Guo+11_sims,
  Guo+13_sims}), we can expect to find $\sim 0.3-3.5$ relic per square
degree, at redshift $z <0.5$.  This prediction does critically depend
on the physical processes shaping size and mass evolution of galaxies,
such as the relative importance of major and minor galaxy merging.

Several compact galaxy definitions have been adopted in the
literature (\citealt{Trujillo+09_superdense};
\citealt{Taylor+10_compacts}; \citealt{Poggianti+13_low_z};
\citealt{Damjanov+15_compacts}). In the present work, we present
the properties of a sample of dense massive galaxy candidates in
KiDS, defining as massive super-compact galaxies (\MSCGs) those
early-type systems with $\mst > 8 \times 10^{10}\, \Msun$ and $\Re
< 1.5 \, \rm kpc$ (\citealt{Trujillo+09_superdense}). These
selection criteria are rather conservative, providing the ideal
benchmark for galaxy evolution theories. The paper is organized as
follows. In \Sec\ref{sec:sample} we present the KiDS data sample
and the selection of our photometrically selected compact
galaxies. The main results, and in particular the evolution of
number density as a function of redshift, are presented in
\Sec\ref{sec:results}.  A discussion of the results and future
prospects are outlined in \Sec\ref{sec:conclusions}. We adopt a
cosmological model with
$(\Omega_{m},\Omega_{\Lambda},h)=(0.3,0.7,0.70)$, where $h =
H_{0}/100 \, \textrm{km} \, \textrm{s}^{-1} \, \textrm{Mpc}^{-1}$
(\citealt{Komatsu+11_WMAP7}).

\section{Sample selection}\label{sec:sample}

The galaxy sample presented in this work is based on the data
included in the first and second data releases of KiDS presented
in \cite{deJong+15_KiDS_paperI}, which we address the interested
reader for details. The total dataset includes 156 KiDS pointings
(133 from the KiDS data release 2), in which we have identified
about 22 million sources including $\sim$7 millions which have
been classified as high quality extended sources (mostly
galaxy-like, see below). A full description of the galaxy sample
is given in Napolitano et al. (2016, in prep.). In the following
section, we summarise the main steps for the galaxy selection
procedure and the determination of galaxy physical quantities as
structural parameters, photometric redshifts and stellar masses.

\subsection{KiDS high signal-to-noise galaxy sample}

We start from the KiDS multi-band source catalogs, where the
photometry has been obtained with S-Extractor
(\citealt{Bertin_Arnouts96_SEx}) in dual image mode, using as
reference the positions of the sources detected in the \rband-band
images. While magnitudes are measured for all of the filters, the
star/galaxy separation, positional and shape parameters are based
on the \rband-band data. The choice of \rband-band is motivated by
the fact that it typically has the best image quality and thus
provides the most reliable source positions and shapes. Critical
areas as saturated pixels, star spikes, reflection halos,
satellite tracks, etc. have been masked using both a dedicated
automatic procedure and visual inspection. The total area after
removing the borders of the individual KiDS pointings is 163.9
square degrees, while the unmasked effective area adopted is of
105.4 square degrees.

Star/galaxy separation is based on the distribution of the
S-Extractor parameters {\tt CLASS\_STAR} and \SN\ (signal-to-
noise ratio) of a number of sure stars (see
\citealt{LaBarbera_08_2DPHOT} and \citealt{deJong+15_KiDS_paperI}
for further details). We have further retained those sources which
were marked as being out of critical area from our masking
procedure. From the original catalog, the star/galaxy separation
leaves $\sim 11$ million galaxies, of which $\sim 7$ millions have
high quality photometry being non-deblended sources located out of
the masked area.

To perform accurate surface photometry and determine reliable
structural parameters, the highest-quality sources have been
further selected (\citealt{LaBarbera_08_2DPHOT, SPIDER-I}). Thus,
we have finally gathered those systems with the highest \SN\ in
the \rband-band images, $\SN_r \equiv$ 1/\MErrautor $> 50$
(\citealt{LaBarbera_08_2DPHOT, SPIDER-I}; Roy et al. 2016, in
preparation). This sample consists of $\sim 380\,000$ galaxies.

Relevant properties for each galaxy are derived as described here below:
\begin{itemize}
\item {\it Photometry.} As a standard KiDS catalogs' parameters, we have derived
S-Extractor aperture photometry in the four bands ($ugri$) within
several radii. For our analysis we have adopted aperture
magnitudes \magaptwo, \magapfour\ and \magapsix, measured within
circular apertures of $2''$, $4''$ and $6''$ of diameter,
respectively, while a first probe of the total magnitude is
provided by the Kron-like \Mauto.
\item {\it Structural parameters.} Surface photometry is
performed using the 2DPHOT environment. 2DPHOT is an automatic
software designed to obtain both integrated and surface photometry
of galaxies in wide-field images. The software first produces a
local PSF model from a series of identified {\it sure stars}. This
is done, for each galaxy, by fitting the four closest stars to
that galaxy with a sum of three two-dimensional Moffat functions.
Then galaxy snapshots are fitted with PSF-convolved S\'ersic
models having elliptical isophotes plus a local background value
(see \citealt{LaBarbera_08_2DPHOT} for further details). The fit
provides the following parameters for the four wavebands: surface
brightness $\mu_{\rm e}$, circularized effective radius, \Re,
S\'ersic index, $n$, total magnitude, $m_{S}$, axis ratio, $q$,
and position angle. As it is common use in the literature, in the
paper we use the circularized effective radius, \Re, defined as
$\Re = \sqrt{q} \Remaj$, where \Remaj\ is the major-axis effective
radius.
\item {\it Photometric redshifts.} Redshifts are determined with the Multi Layer
Perceptron with Quasi Newton Algorithm (MLPQNA) method
(\citealt{Brescia+13, Brescia+14}), and fully presented in
\cite{Cavuoti+15_KIDS_I}, which we refer for all details. Both
apertures of $4''$ and $6''$ of diameter are used. In machine
learning supervised methods, a knowledge sample is needed to train
the neural network performing the mapping between magnitudes and
redshift. The knowledge base consisted on spectroscopic redshift
from the Sloan Digital Sky Survey data release 9 (SDSS-DR9;
\citealt{Ahn+12_SDSS_DR9}) and Galaxy And Mass Assembly data
release 2 (GAMA-DR2; \citealt{Driver+11_GAMA}) which together
provide redshifts up to $z \lsim 0.8$. This knowledge base
includes $\sim 60000$ objects, 60\% of which are used as training
sample, and the remaining ones are used for the blind test set.
The redshifts in the blind test sample resemble the spectroscopic
redshifts quite well. The scatter in the quantity $\Delta z \equiv
(\zspec - \zphot)/(1+\zspec)$ is $\sim 0.03$. After these
experiments, we have finally produced the final catalogue of
redshifts for our sample. The cut operated in the fitting
magnitudes to resemble the luminosity ranges in the knowledge base
will impact the completeness in the faint regime. But our
high-\SN\ sample is not affected by this source of incompleteness.
\item {\it Stellar masses.} We have used the software {\tt Le Phare}
(\citealt{Arnouts+99}; \citealt{Ilbert+06}), which performs a
simple $\chi^{2}$ fitting method between the stellar population
synthesis (SPS) theoretical models and data. Single burst models
from \cite{BC03} and a \cite{Chabrier01} IMF are used. Models are
redshifted using the photometric redshifts. We adopt the observed
$ugri$ magnitudes (and related $1\, \sigma$ uncertainties) within
a $6''$ aperture of diameter, which are corrected for Galactic
extinction using the map in \cite{Schlafly_Finkbeiner11}. Total
magnitudes derived from the S\'ersic fitting, $m_{S}$, are used to
correct the outcomes of {\tt Le Phare} for missing flux. The
single burst assumption is suitable to describe the old stellar
populations in the compact galaxies we are interested in
(\citealt{Thomas+05}; \citealt{Tortora+09}). The estimated
photometric ages are used to check if galaxies are compatible with
being relic remnants of systems formed at $z\sim2$. The degeneracy
between age and metallicity in the stellar population analysis can
be solved only with forthcoming spectroscopic follow-ups.
\item {\it "Galaxy classification".} Using {\tt Le Phare}, we have
also fitted the observed magnitudes \magapsix\ with a set of $66$
empirical spectral templates used in \cite{Ilbert+06}. The set is
based on the four basic templates (Ell, Sbc, Scd, Irr) described
in \cite{CWW80}, and star burst models from \cite{Kinney+96}.
GISSEL synthetic models \citep{BC03} are used to linearly
extrapolate this set of templates into ultraviolet and
near-infrared. The final set of $66$ templates ($22$ for
ellipticals, $17$ for Sbc, $12$ for Scd, $11$ for Im, and $4$ for
starburst) is obtained by linearly interpolating the original
templates, in order to improve the sampling of the colour space.
We have selected the ETGs by choosing the galaxies which are best
fitted by one of the elliptical templates (see Napolitano et al.
2016, in preparation).
\item {\it Near Infrared photometry from VIKING-DR1.}
As a complementary dataset for our selection of compact galaxies,
we have used the data from the first release (DR1) of the ESO
VISTA Kilo Degree Infrared Galaxy (VIKING) survey. The VIKING
survey is the KiDS twin survey, and provides the near-IR coverage
of the same sky region in the 5 wavebands Z, Y, J, H and Ks.
VIKING-DR1 consists of 108 observed tiles
(\citealt{Edge+14_VIKING-DR1}). We have found that it overlaps
with $\sim 58 \%$ of our 156~sq.~deg. in the KiDS survey. For
uniformity, we have used photometry within the same aperture as
the optical data: \amthree\ for VIKING and \magaptwo\ for KiDS,
which correspond to an aperture of $2''$ of diameter. In
particular, we have concentrated on J and Ks passbands, which we
do not use for SPS fitting, but only to improve the star-galaxy
separation criterion (\citealt{Maddox+08}; \citealt{Muzzin+13}) as
will be discussed in details in the next section.

\end{itemize}

\begin{figure}
\centering \psfig{file=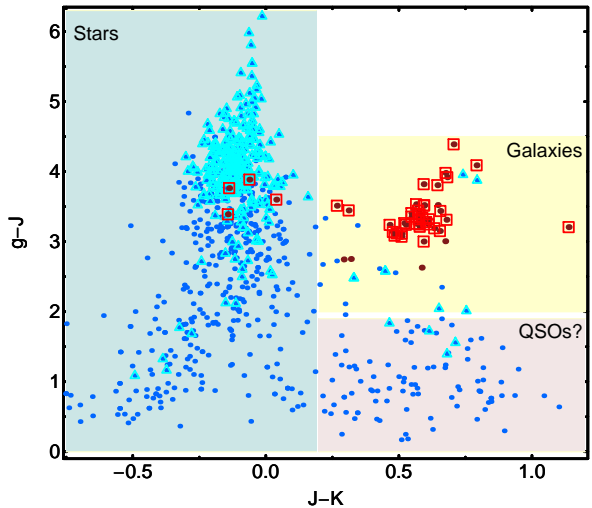, width=0.45\textwidth}
\caption{$J-K$ vs. $g-J$ diagram. An aperture of $2''$ of diameter
is adopted here. Blue and cyan symbols are for high-confidence
stars with any error and  $\delta J = \delta K < 0.05$,
respectively. Red points and open boxes are for compact candidates
with any error and $\delta J = \delta K < 0.05$, survived after
the criteria on the mass, size, 2D fit quality and visual
inspection. Vega J and K magnitudes are converted to AB as $J_{\rm
AB}= J_{\rm Vega}+ 0.930$ and $K_{\rm AB}= K_{\rm Vega}+ 1.834$.
We highlight the regions which are populated by stars (blue),
galaxies (yellow) and QSOs (pink). We have considered as sure
galaxies those objects with colours $J-K> 0.2$ and $g-j
> 2$ (yellow shaded region). }\label{fig:colour-colour}
\end{figure}

The sample of high-\SN\ galaxies is complete down to a magnitude
of  $\Mautor \sim 21$, which correspond to stellar masses $\gsim 5
\times 10^{10}$ up to redshift $z = 0.5$ (see see Napolitano et al. 2016,
in prep., for further details).

\begin{figure*}
\centering \psfig{file=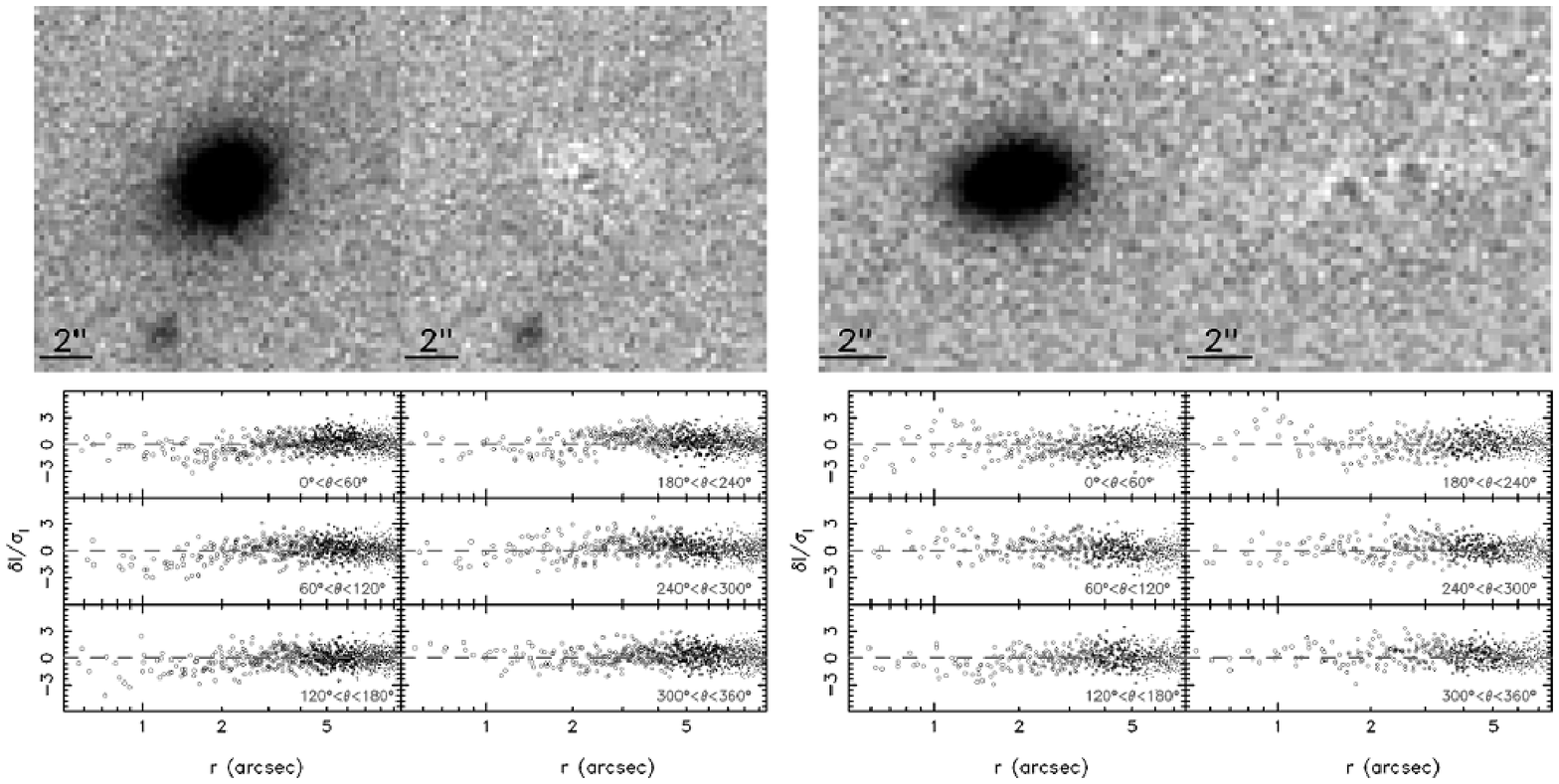,width=0.49\textwidth}
\psfig{file=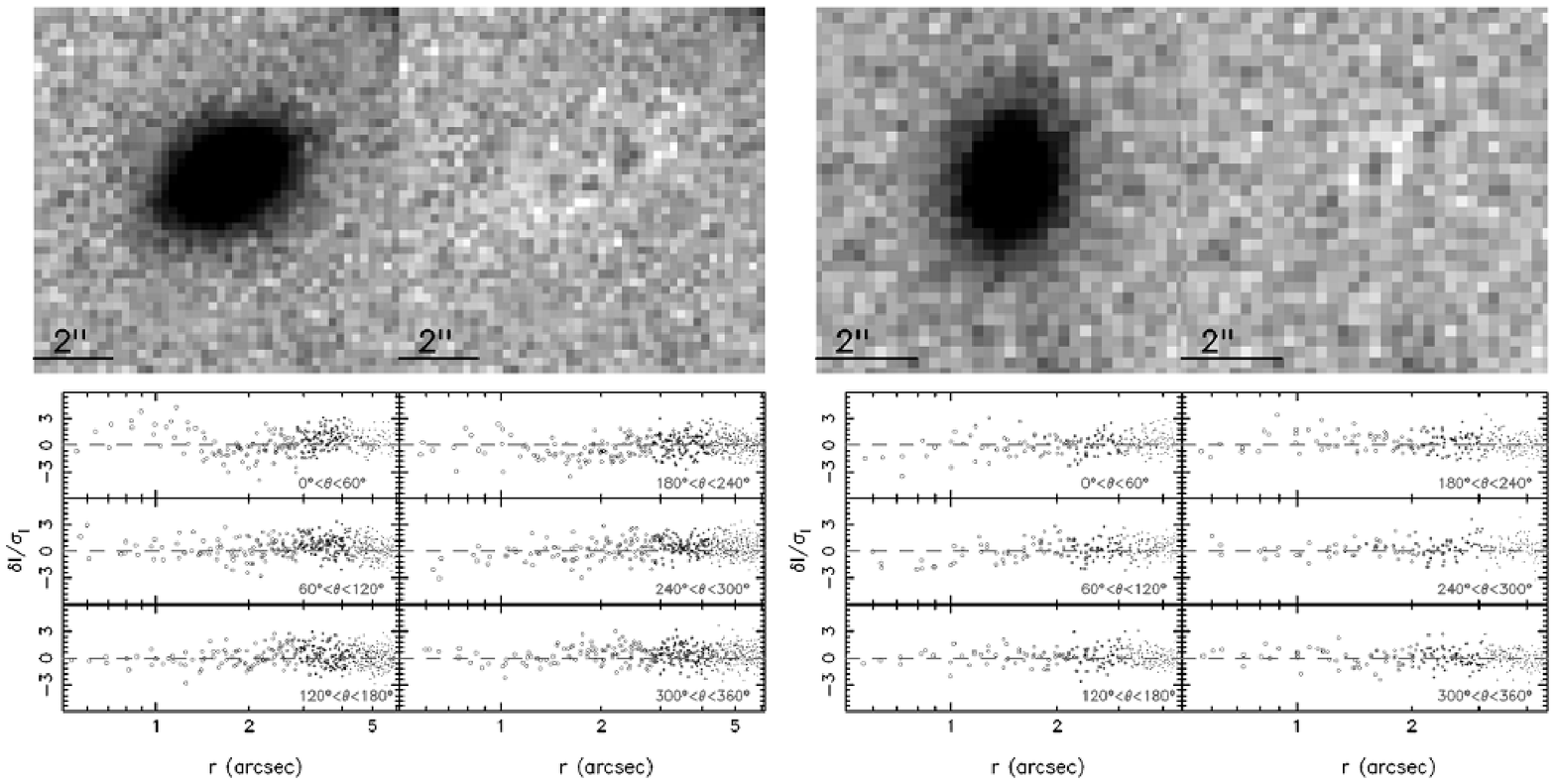,width=0.49\textwidth} \caption{2D fit output
for 4 example candidates from 2DPHOT procedure. For each galaxy,
the top panels show the galaxy image (left) and the residual after
the fit (right), while the six bottom panels provide residuals
after subtraction, plotted as a function of the distance to the
galaxy center, with each panel corresponding to a different bin of
the polar angle. Residuals are normalized with the noise expected
from the model in each pixel.}\label{fig:2D}
\end{figure*}

\subsection{Selection of compact galaxies}

\MSCGs\ have been selected using the following criteria:
\begin{enumerate}
\item {\it Massiveness}. The most massive galaxies
with $\mst
> 8 \times 10^{10}\, \rm \Msun$ are taken
(\citealt{Trujillo+09_superdense}), reducing the original sample
of $\sim 380000$ galaxies to $\sim 30600$ massive galaxies.
\item {\it Compactness}. We select the densest galaxies by following recent
literature (\citealt{Trujillo+09_superdense}). We get galaxies
with median circularized radius, $\Re$, among the  \g-, \rband-
and \i-bands, less than $ 1.5 \, \rm kpc$. About 1300 compact
candidates remain after this selection.
\item {\it Best-fitted effective radii}. The best fitted structural parameters
are considered, taking those systems with a reduced
$\chi^{2}_{2D}$ from {\tt 2DPHOT} smaller than $1.5$ in \g,
\rband\ and \i\ filters (\citealt{SPIDER-I}). To avoid any
accidental wrong fits we have also removed galaxies with
unreasonable best-fitted parameters, applying a minimum value for
the size ($\Re = 0.05$ arcsec), the S\'ersic index ($n>0.5$) and
the axial ratio ($q = 0.1$) in all the bands. The minimum value in
the S\'ersic index has also allowed us to possibly remove
misclassified stars, which are expected to be fitted by a box-like
profile (mimicked by a S\'ersic profile with $n \to 0$). These
last criteria on the quality of the structural parameters reduce
further the sample to the 106 highly reliable candidates.
\item {\it Eye-ball check}. We have made an eyeball inspection of the
  images and residuals from S\'ersic fitting of the candidates,
  with the aim of removing problematic objects or possible
  misclassified stars.  To reduce subjectivity, three of the authors
  have independently checked the images and graded them
  according to the following scheme: \gtwo\ are sure galaxies,
  \gone\ are uncertain galaxies lacking a well-defined elliptical
    shape, and \gzero\ are misclassified objects (either stars or
  corrupted fits).  The mean of the three classifications has been
  adopted as the final grade.  In this way, we have retained
  those galaxies with a grade larger than $1$, to include only systems
  graded by at least one of the observers  as a sure
  galaxy.   The candidates are further reduced to 96, after
    removing objects with significant contamination from
    neighbours.

\item {\it Optical+NIR star-galaxy separation.} We have adopted a
  morphological criterion to perform the star-galaxy classification
  (\citealt{Bertin_Arnouts96_SEx}; \citealt{LaBarbera_08_2DPHOT}) and
  used visual inspection as ultimate check of the galaxy
  classification. However, based on optical data only a star can be
  still misclassified as a galaxy on the basis of its morphology, and
  this issue can be highly dramatic for very compact objects
  (generally with size comparable or smaller than the seeing).  In
  absence of spectroscopic information, optical+NIR colour-colour
  diagrams can provide a strong constraint on the galaxy nature of the
  candidates (e.g.  \citealt{Maddox+08}; \citealt{Muzzin+13}).  In
  particular, \g, \J\ and \Ks\ magnitudes within $2''$ of diameter are
  adopted for this purpose on those fields with coverage by the two
  surveys. The Vega VISTA magnitudes are converted to AB using the
  conversion formulae from {\tt Le Phare}, $J_{\rm AB}= J_{\rm Vega}+ 0.930$ and $K_{\rm
    AB}= K_{\rm Vega}+ 1.834$. Stars and galaxies with the best J and
  Ks photometry are also considered ($\delta J, \, \delta K <
  0.05$). The results are shown in \Fig\ref{fig:colour-colour}. Stars
  have blue $J-K$ colours (i.e., $J-K \lsim 0.2$, see light blue
  shaded region in \Fig\ref{fig:colour-colour}).  We have also found
  four objects with $J-K \lsim 0.2$ and, indeed, are erroneously
  classified as galaxies. We take as compact candidates those systems
  with $J-K> 0.2$ and $g-J > 2$ (see light-yellow shaded region in
  \Fig\ref{fig:colour-colour}).  For those galaxies with available
  VIKING photometry, this selection allows our set of candidates
    from the previous classification steps to be refined.  Out of
    46 galaxies with VIKING photometry, we remain with 42
    high-confidence candidates.

In \Fig\ref{fig:colour-colour} we have also highlighted the locus
populated by point sources with red $J-K$ ($\gsim 0.2$), but blue
$g-J$ ($\lsim 2$) colours, classified as stars by S-Extractor,
which are presumably quasars (\citealt{Maddox+08}, see pink region
in \Fig\ref{fig:colour-colour}). However, the analysis of this
class of objects is beyond the scope of the paper.
\end{enumerate}

\begin{figure*}
\centering \psfig{file=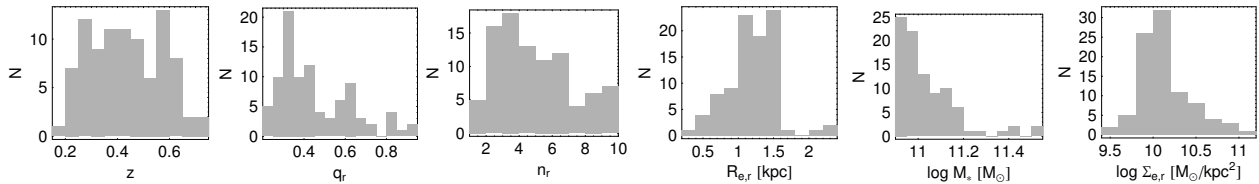, width=0.99\textwidth}
\caption{Distribution of some galaxy parameters. From left to
right we show: 1) photometric redshifts, 2) r-band axis ratio, 3)
r-band S\'ersic index, 4) r-band effective radius, 5) Chabrier
IMF-based stellar mass, 6) r-band effective surface mass
density.}\label{fig:hist}
\end{figure*}

To  perform an  homogeneous comparison  of the  sample of  our
compact candidates with a sample of "normal" galaxies, using the
same original catalog described in \Sec\ref{sec:sample}, we  have
created two control samples (\CSs) of galaxies  with similar
stellar masses ($\mst
> 8  \times 10^{10}  \, \rm  \Msun$), the same  lower limits  on
  S\'ersic index  ($n > 0.5$), axis  ratio ($q>0.1$) and \Re\  ($\Re >
  0.05$ arcsec) as those for the compact systems, but without applying
  any compactness  criterion. The  first \FULL\  \CS\ consists  of all
  galaxies,  with  no  restriction  on  the  galaxy  type,  while  the
  \ETGs\   \CS\  consists   of  ETGs,   classified  as   described  in
  \Sec\ref{sec:sample}.

\section{Results}\label{sec:results}

\subsection{The final sample}

We start by summarizing the sample we are left with for our
analysis: after the first three criteria we have selected 106
candidates, which are reduced to 96 after the eye-ball check in
(iv). The matching with VIKING included 46 candidates and, after
applying the criterion (v), we are left with a sample of 42 out 46
galaxy candidates (i.e. 4/46 objects are likely stars according to
their optical-NIR colours, corresponding to a $\sim10\%$
contamination).  Updating the total number, we are left with 92
candidates, of which $\sim 4$ further sources from the region that
does not overlap with VIKING might be stars. This corresponds to a
density of $\sim 0.9$ compact galaxies per square degree,
cumulatively for $z\lsim0.7$, while we do not find objects at
$z\lsim0.2$. In contrast, \citet{Trujillo+09_superdense} find 29
secure \MSCGs\ at $z < 0.2$ within the SDSS area, using our same
criteria. However, only one of these is at very low redshift
($z<0.1$), and none of them features old stellar populations
(\citealt{Ferre-Mateu+12}), and thus they cannot actually be
classified as relics. The absence of compact systems at $z<0.2$ in
our datasample might be related to environmental issues, since in
the current KiDS area we are missing nearby clusters of galaxies,
where a larger fraction of compact galaxies is predicted
(\citealt{Stringer+15_compacts}), and seems to be observed indeed
(\citealt{Poggianti+13_low_z}; \citealt{Valentinuzzi+10_WINGS}). A
similar environment effect would be present in the
\cite{Trujillo+09_superdense} sample. Examples of the 2D fitting
results for four candidates are presented in \Fig\ref{fig:2D},
where postage stamps of the galaxy images and the residuals are
shown.

\Fig\ref{fig:hist} shows the distributions of redshift, axis
ratio, S\'ersic index, effective radius, stellar mass and surface
mass density for our sample of compact candidates. The median
redshift of the sample is $\sim 0.44$, with an rms of $\sim 0.18$,
larger than the median redshifts of the \FULL\ and \ETGs\ \CSs\
($\sim 0.36$ and $\sim 0.34$, respectively). Compact candidates
have a median r-band effective radius of $\sim 1.2$ kpc (rms=0.31
kpc), a S\'ersic index of $\sim 4.3$ (rms=2.3), an axis ratio of
0.40 (rms=0.14) and median stellar mass of $\sim 10.99$ dex
(rms=0.09). On the contrary, for the \FULL\ \CS\, the median size,
S\'ersic index, axis ratio and \mst\ are $\sim 7.7$ kpc, 4.7, 0.73
and $\sim 11.07$ dex, respectively. Finally, if the \ETGs\ \CS\ is
considered, the median size, S\'ersic index, axis ratio and \mst\
are $\sim 8.4$ kpc, 5.5, 0.75 and $\sim 11.10$ dex, respectively.
Thus, smaller sizes translate into shallower light profiles
(smaller $n$) when compacts are compared with normal galaxies
(consistently with the galaxy merging framework in
\citealt{Hilz+13}). Compact candidates also present lower axis
ratios than normal galaxies. These elongated shapes are common in
massive compact galaxies, both at high-z (\citealt{vanderWel+11};
\citealt{Buitrago+13}), and at lower redshifts
(\citealt{Trujillo+12_compacts, Trujillo+14}).

In our compact sample, the average surface mass density within 1
\Re\ is $\sim 1.2 \times 10^{10} \, \rm \Msun / kpc^{2}$ (rms of
$\sim 5.6 \times 10^{9} \, \rm \Msun / kpc^{2}$), about 2 orders
of magnitude larger than the average surface mass density of the
\CSs . This median density is a bit larger than the range of
values of $7-15 \times 10^{9} \, \rm \Msun / kpc^{2}$ found in 4
compact galaxies at $z<0.2$ by \cite{Trujillo+12_compacts}, but
the values are consistent within the scatter. Taking the only 3
galaxies from \cite{Szomoru+12}, which satisfy the same criteria
we are adopting in this paper, we find a median density of $\sim
1.1 \times 10^{10} \, \rm \Msun / kpc^{2}$, which is fully
consistent with our result.

The \rband-band \Re\ as a function of \mst\ is shown in \Fig
\ref{fig:size_mass}. The compact candidates are plotted with a
random subset of $\sim 2500$ galaxies from the \FULL\ \CS. As the
compact candidates are selected to have small sizes, they lie in a
region of the size-mass diagram where very few objects are
detected, which provides a further evidence in favor of the rarity
of these compact objects at $z \lsim 0.7$
(\Fig\ref{fig:size_mass}; \citealt{Trujillo+09_superdense}). In
\Fig \ref{fig:size_mass} we also overplot the $z<0.2$ galaxies in
\cite{Trujillo+12_compacts} and the three $z \sim 2$ galaxies in
\cite{Szomoru+12} which fullfil our selection criteria. The
evolution with redshift of the sizes, magnitudes and colours are
shown in \Fig\ref{fig:vari_vs_z}. The \Re s look quite constant as
a function of redshift for the compact systems. At all the
redshifts, almost all the compacts have faint \Mauto\ when
compared with the control sample population (middle panel in
\Fig\ref{fig:vari_vs_z}). We find that 75 candidates ($\sim 82\%$
of the whole sample) lie below the r-band magnitude completeness
limit of $\sim 21$, implying that our sample is complete up to
redshift $z\sim 0.5$. Finally, most of the galaxies populate the
red-sequence, and are the best candidates to be remnants of high-z
red nuggets (right panel in \Fig\ref{fig:vari_vs_z}).

58 out of 92 galaxies have old ages consistent with a formation
redshift $z_{\rm f} \gsim 2$, and so could be the remnants of the
compact galaxies observed at $z > 1$ (\citealt{Gargiulo+11,
Gargiulo+12}; \citealt{Szomoru+12}). 82 out of 92 galaxies (i.e.
$\sim 89\%$ of the whole sample) are best fitted by an elliptical
template, and are classified as ETGs (see \Sec \ref{sec:sample}).
The rest of the galaxies have redshifts $z>0.5$ and have colours
which are best-fitted by Sbc models. However, particularly at
these redshifts, one should take this colour-based classification
with caution, as a spectroscopic follow-up is actually needed to
perform an accurate stellar population analysis.

\subsection{Systematics from wrong
redshifts}\label{subsec:wrong_z}

We have finally cross-matched our sample of candidates with SDSS
and GAMA spectroscopy catalogs, finding spectra for 9 of the
selected candidates.

\begin{figure}
\centering \psfig{file=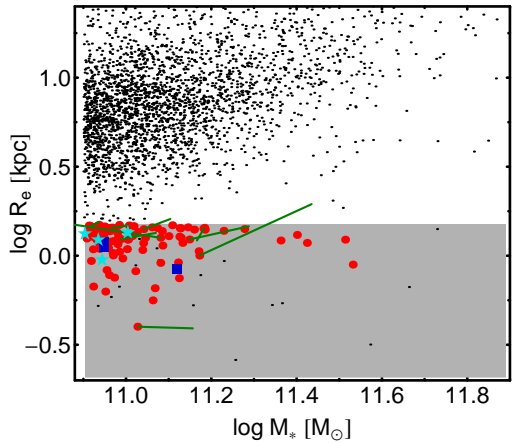, width=0.4\textwidth} \caption{We
show the  median of the \Re\ in the g-, r- and i- bands vs.
stellar mass. Compact candidates (plotted as red points) are
compared with a selection of galaxies ($\sim 2500$ randomly
extracted) from the \FULL\ \CS\ (small black points). Gray shaded
region highlight the region where candidates are selected (see
text for details about the adopted selection criteria). We also
plot the change if the spectroscopic redshift is used (dark-green
lines). Finally, the four compact galaxies with $z < 0.2$ from
\citet{Trujillo+12_compacts} and the three compacts at $z \sim 2$
from \citet{Szomoru+12}, that fulfil our selection criteria, are
shown as cyan stars and blue squares,
respectively.}\label{fig:size_mass}
\end{figure}

\begin{figure*}
\centering \psfig{file=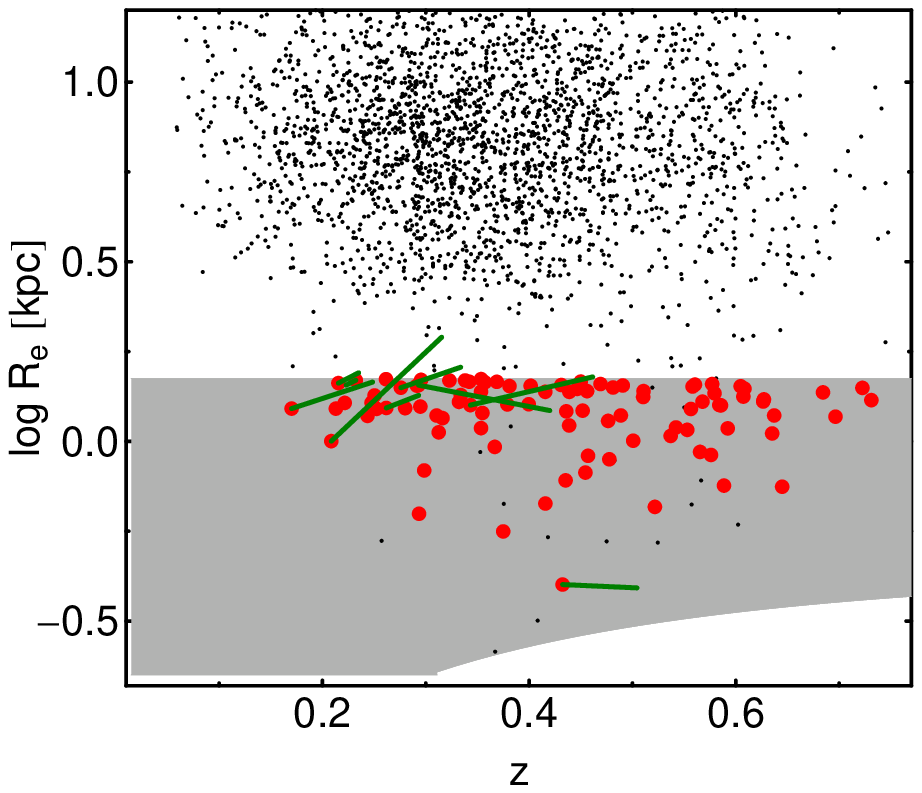, width=0.32\textwidth}
\psfig{file=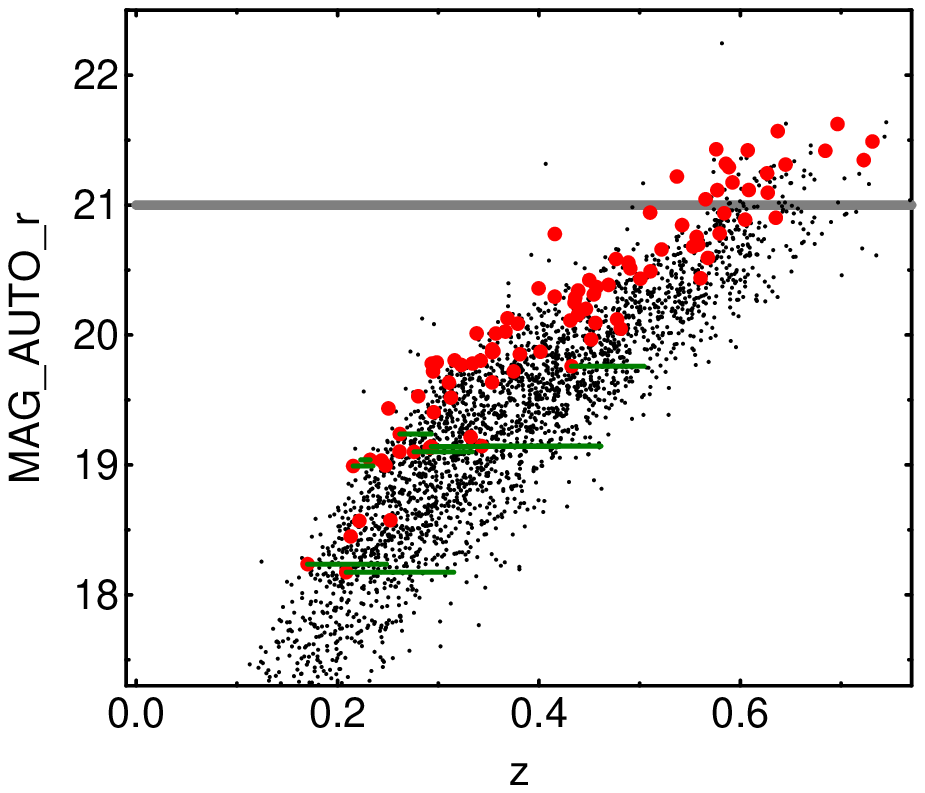, width=0.32\textwidth}
\psfig{file=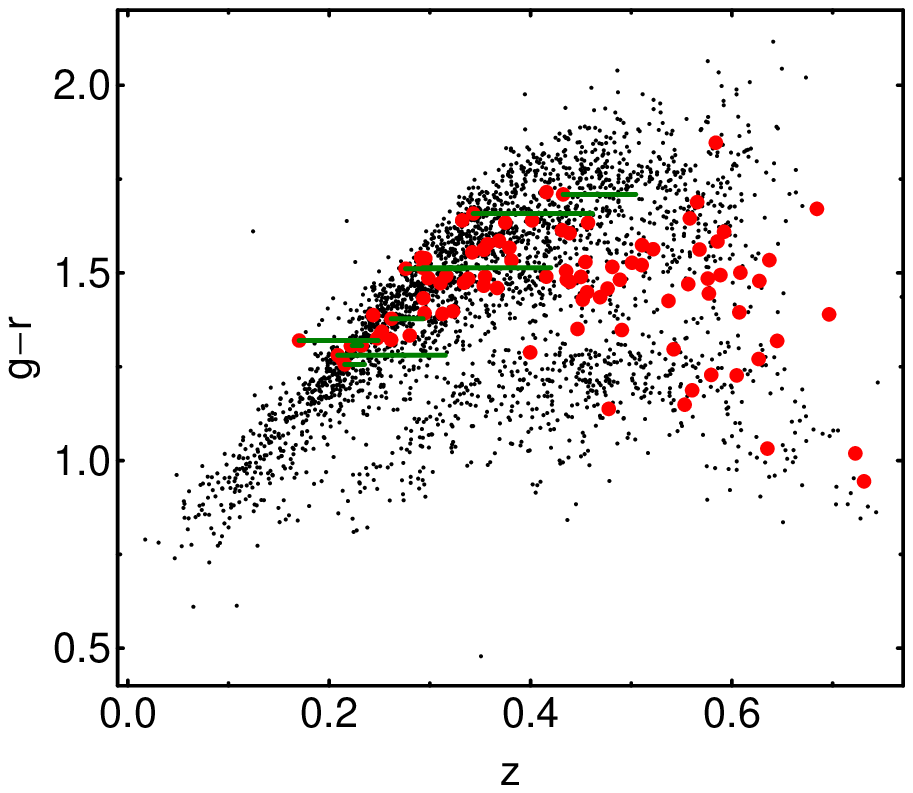, width=0.32\textwidth} \caption{We show the
median of the \Re\ in the g-, r- and i- bands (left), r-band
\Mauto\ (middle) and $g-r$ (right) vs. redshift.  The gray region
in the left panel sets the region within which we have searched
the compacts. Gray line in the middle panel sets the 90\%
completeness limit of the high--\SN\ sample. The $g-r$ colour is
calculated within $6''$ of diameter. See further details about the
symbols in \Fig \ref{fig:size_mass}.}\label{fig:vari_vs_z}
\end{figure*}

One of the main systematics in our selection of compact galaxies
is induced by wrong redshift determinations, which can affect both
the (linear) effective radii and stellar masses, moving the
compact out of our selection criteria (e.g., see \Figs
\ref{fig:size_mass} and \ref{fig:vari_vs_z}). 7 out 9 of these
systems are included in the SDSS+GAMA datasample used for
photometric redshift determinations (see \Sec \ref{sec:sample}): 5
in the training sample and 2 in the test sample. Although the
photometric redshifts are shown to approximate quite well the
spectroscopic ones (see \Sec \ref{sec:sample} and
\citealt{Cavuoti+15_KIDS_I}), we note that almost all of the
photometric redshifts are underestimated with respect to the
spectroscopic value. The median difference $\Delta z \equiv \zphot
- \zspec$ is $-0.07$, which increases to $-0.1$ if the galaxies
not in the training sample are considered. This systematics can be
probably related to the under-sampling of this galaxy population,
just 5 galaxies in the training sample, which can fail the optimum
training of the network, and the quality of photometric redshift
outcomes is degraded.

As first test, we have re-computed the sizes and stellar masses of
these 9 objects using the \zs . Only 4 galaxies survive to the
size and mass selection criteria, i.e. $\sim 44\%$ (1 not in the
training sample).

Despite the small sample available for this test, we cannot
exclude the presence of a systematics of the photometric redshift
determination for compact sources which we might quantify of the
order of $\Delta z = - 0.1$. Qualitatively, if the spectroscopic
redshift is larger than photometric one, then \Re\ in physical
scale gets larger. The effect on \mst\ is not as simple as the one
on \Re. As an exercise, we have investigated the impact of a
redshift error of $\Delta z=0.1$ on the 83 galaxies in the sample
without spectroscopic redshift. Using re-calculated values for
\Re\ and stellar mass, we find that 26 of the sources (31\%)
survive the cuts.

Therefore, a spectroscopic follow-up will be needed to finally
confirm the nature of these galaxies, allowing us to classify
these candidates as compact or very-compact galaxies. Increasing
the knowledge base and the population of compact systems with
measured spectroscopic redshifts will improve the quality of the
photometric redshift estimates.

\subsection{Colour gradients}

A big improvement of our analysis with respect to previous works
on compact galaxies at similar redshift ranges (e.g.
\citealt{Damjanov+14_compacts}) is the high-\SN\ photometry which
allows us to derive robust structural parameters and obtain the
first determination of colour profiles for these systems.
Therefore, following \cite{Tortora+10CG}, we define the colour
gradient as the angular coefficient of the relation $X-Y$ vs.
$\log R$, $\displaystyle \nabla_{X-Y} = \frac{\delta (X-Y)}{\delta
\log R}$, measured in $\rm mag/dex$ (omitted in the following
unless needed for clarity). The fit of each color profile is
performed in the range $R_{1}= 0.1 \Re \leq R \leq R_{2}= \Re$,
where the effective radius is measured in the \rband-band. By
definition, a positive CG, $\nabla_{X-Y}>0$, means that a galaxy
is redder as $R$ increases, while it is bluer outward for a
negative gradient. PSF-convolved S\'ersic profiles in \g-, \rband-
and \i-band are used and, in particular, $g-r$, $r-i$ and $g-i$
colours are discussed. We omit detailed analysis in terms of
redshift and stellar mass, but we pinpoint what is the range of
colour gradients in our galaxy sample, comparing these results
with the \ETGs\ \CS\ and some previous literature.

On average, compact population has $\ggr = -0.21$ (rms 0.52),
$\gri = -0.07$ (rms 0.59) and $\ggi = -0.30$ (rms 0.73) which are
substantially consistent with the gradients for the control
population of ETGs which are $\ggr = -0.17$ (rms 0.33), $\gri =
-0.05$ (rms 0.23) and $\ggi = -0.21$ (rms 0.40). Hence, compact
candidates look quite similar to normal ETGs, except for the
scatter, which is partially or totally related to the larger
uncertainties on the structural parameters in our small objects.

These results agree with previously reported ranges of color
gradients of passively evolving massive galaxies at low- or
intermediate-z (\citealt{Tamura+00}; \citealt{Tamura_Ohta00};
\citealt{Wu+05}; \citealt{Tortora+10CG};
\citealt{LaBarbera+12_SPIDERVII_CG}) or at high-z
(\citealt{Guo+11}; \citealt{Gargiulo+11,Gargiulo+12}) and with
simulations (e.g., \citealt{Tortora+13_CG_SIM}).

Finally, we find that 30 compact candidates have all negative
colour gradients, 9 have positive gradients, while the rest of the
sample have at least one of the three gradients with a different
sign with respect to the others. This wide range of behaviours
demonstrate that these objects can be formed in quite different
initial conditions (\citealt{Gargiulo+12};
\citealt{Damjanov+14_compacts}).

\begin{figure*}
\centering \psfig{file=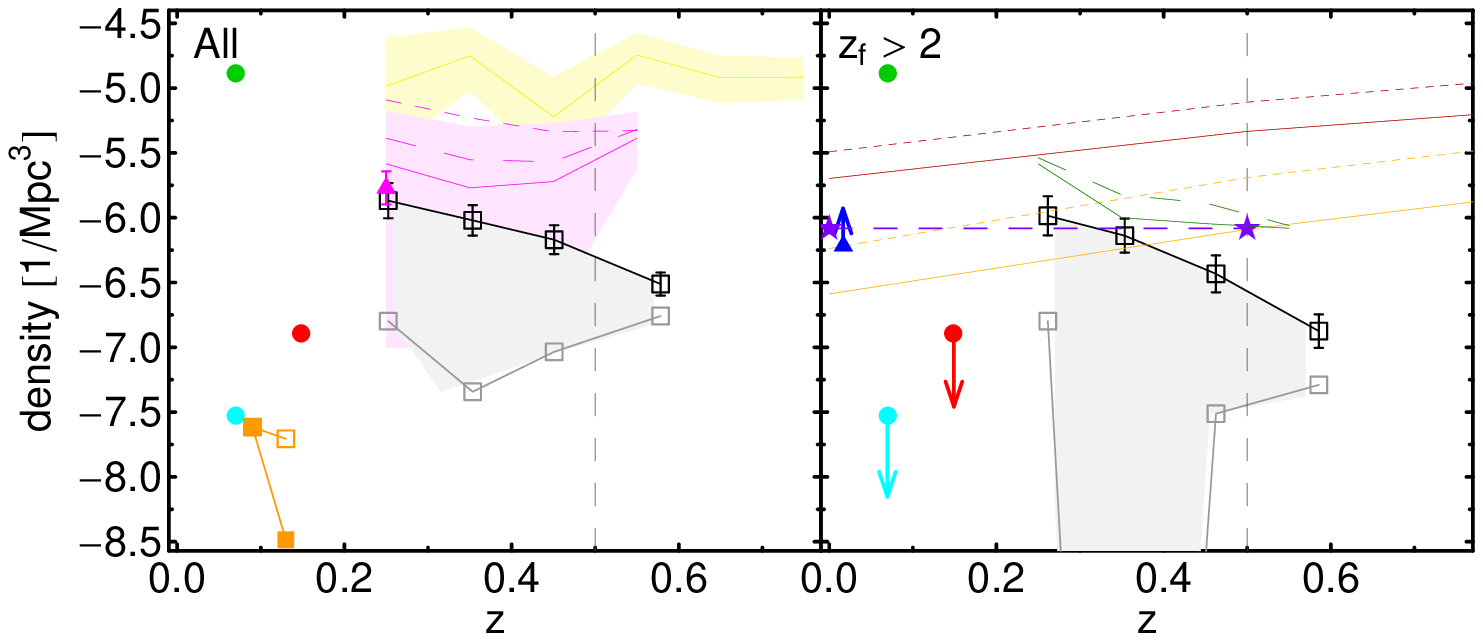, width=0.68\textwidth}
\psfig{file=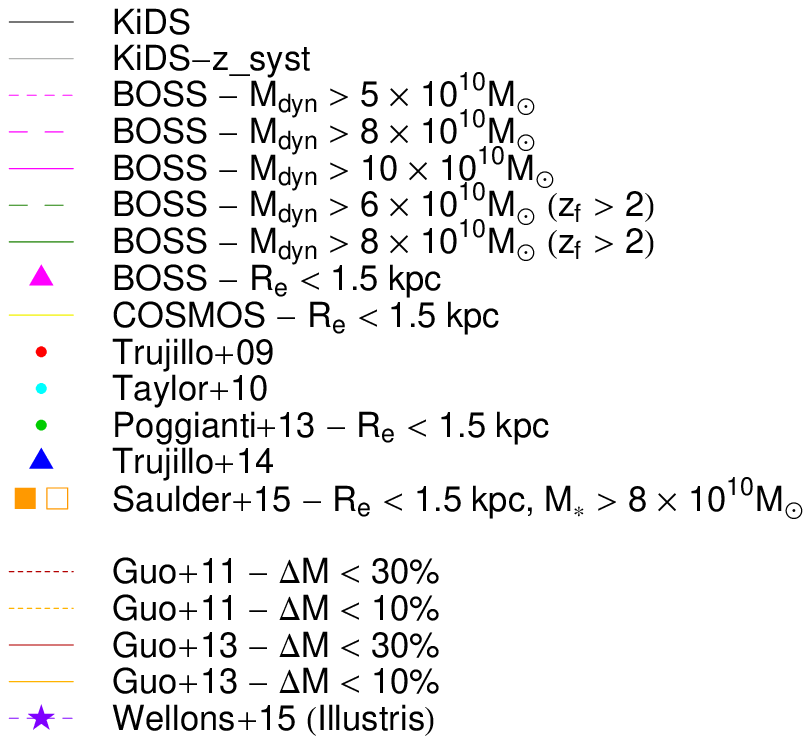, width=0.26\textwidth, bb=60 -40 300 100}
\caption{Number density of compact galaxies vs.  redshift. Error
bars denote $1\sigma$ uncertainties (see the text). Vertical
dashed black lines mark the completeness limit of the sample (see
\citealt{deJong+15_KiDS_paperI}).  {\it Left.}  Number densities
for all galaxies, independent of their photometric formation
redshift, are plotted with open squares and solid black lines. The
solid gray line with open squares takes into account possible
systematics in the redshift determination (see
\Sec\ref{subsec:wrong_z}).  Short-dashed, long-dashed and solid
violet lines are number densities of stellar-like objects from
BOSS-DR10 (\citealt{Damjanov+14_compacts}) with $\Mdyn > 5 \times
10^{10}$, $8 \times 10^{10}$, and $10^{11}$ \Msun, respectively.
The violet shaded region is the $1\sigma$ error for the case with
$\Mdyn > 10^{11}$~\Msun .  The violet triangle with bars also
shows the abundance of galaxies at $z\sim 0.25$, with $\Re < 1.5\,
\rm kpc$ and $\Mdyn > 8 \times 10^{10}\, \rm \Msun$, from
\citet{Damjanov+14_compacts}.  The yellow line with lighter yellow
region plot abundances for compacts in the COSMOS area
\citep{Damjanov+15_compacts}, selected with the same criteria as
in the present work ($\mst > 8 \times 10^{10} \, \rm \Msun$ and
$\Re < 1.5\, \rm kpc$; I. Damjanov, private communication).  Red,
cyan and green points are the results for compact galaxies from
\citet{Trujillo+09_superdense}, \citet{Taylor+10_compacts} and
\citet{Poggianti+13_low_z}, respectively. Orange boxes show the
abundances for compacts in SDSS area from
\citet{Saulder+15_compacts}, adopting our same criteria on mass
and size. Filled boxes plot the results using S\'ersic profiles,
while open boxes are for the de Vaucouleurs profile (note that the
results for the two profiles in the lowest redshift bin are
superimposed). {\it Right.} Black open squares, solid lines and
error bars plot KiDS number densities for candidate relic
compacts, defined to have photometric formation redshift $z_{f} >
2$. Gray open squares and lines take into account possible
systematics in the redshift determination. Long-dashed and solid
green lines are for stellar-like objects from BOSS-DR10 with
$z_{f}>2$ (\citealt{Damjanov+14_compacts}) and $\Mdyn > 6 \times
10^{10}$ and $8 \times 10^{10}$, respectively. The results from
\citet{Trujillo+09_superdense} and \citet{Taylor+10_compacts} are
shown here as upper limits (see red and cyan points with arrows).
The blue triangle is for the lower limit at $z \sim 0$ provided by
\citet{Trujillo+14}.  Orange and red lines plot abundances of
relic galaxies from semi-analytical models (SAMs), based on
Millennium N-body simulations (\citealt{Quilis_Trujillo13}).
Relics have been defined as systems whose stellar mass has
increased since $z = 2$ to the present by less than $10\%$ and
$30\%$, respectively.  Dashed and solid lines are for
\citet{Guo+11_sims} and \citet{Guo+13_sims} SAMs, respectively.
Purple stars (connected by a line) are predictions from the
Illustris simulations
\citep{Wellons+15_lower_z}.}\label{fig:density_vs_z}
\end{figure*}

\subsection{Abundance vs. redshift}

The number density of compact massive galaxies as a function of
redshift is an important constraint on models of galaxy assembly.
In recent years there have been different efforts to produce a
census of such systems in different redshift bins (e.g.
\citealt{Trujillo+09_superdense}; \citealt{Taylor+10_compacts};
\citealt{Cassata+13}; \citealt{Poggianti+13_evol,
Poggianti+13_low_z}; \citealt{Damjanov+14_compacts,
Damjanov+15_compacts}, \citealt{Saulder+15_compacts}). Sample size
is a crucial aspect to increase the constraining power. If the
compact galaxies found in the present work are a representative
subsample of the whole population of compacts over the whole area
of 1500~sq.~deg. that will be mapped by the KiDS survey, we expect
to increase the present sample by a factor of ten in the next few
years.

For what concerns our current sample, we have binned galaxies with
respect to redshift and normalized to the comoving volume
corresponding to the observed KiDS sky area\footnote{This is
obtained by multiplying the number of candidates by $f_{\rm area}
= A_{\rm sky}/A_{\rm survey}$, where $A_{\rm sky}$
($=41253$~sq.~deg.) is the full sky area and $A_{\rm
survey}=105.4$ is the effective KiDS-DR2 area.  Then, the density
is derived by dividing for the comoving volume corresponding to
each redshift bin.}. The errors on number counts take into account
fluctuations due to Poisson noise, as well as those due to
large-scale structure (i.e. the cosmic variance): they are
calculated with the online {\tt
CosmicVarianceCalculator}\footnote{http://casa.colorado.edu/$\sim$trenti/CosmicVariance.html}
tool (\citealt{Trenti_Stiavelli08}).  The cosmic variance
increases the Poissonian error budget by $\sim 5-30\%$.  The total
relative error on abundances (i.e. number densities) amounts to
$\sim 25-30\%$.

In \Fig\ref{fig:density_vs_z} we plot the redshift evolution of
the abundance of compact galaxies (left panel) and that for the
subsample of systems with old photometric ages (i.e.  formation
redshift $z_{f} \geq 2$; right panel).  We consider these
potentially old \MSCGs, as candidate remnants of the compact ETGs
found at high redshift by several studies (see
\Sec\ref{sec:intro}). We also re-determine the abundances by
accounting for possible systematics in the photometric redshifts
(see \Sec\ref{subsec:wrong_z}; see grey symbols in
\Fig\ref{fig:density_vs_z}). We remind the reader that our sample
starts to be incomplete at $z \gsim 0.5$ (see vertical dashed
lines).

In the left panel of \Fig\ref{fig:density_vs_z}, we plot number
densities for the whole sample of KiDS compacts, independent of
the galaxy formation redshift. Our number densities are compared
with estimates from \cite{Damjanov+14_compacts}, for the number
density of stellar-like objects having spectroscopic redshifts
from BOSS-DR10 (\citealt{Ahn+14_SDSS_DR10}) and with three
different cuts in total dynamical mass, \Mdyn\ ($> 5 \times
10^{10}$, $8 \times 10^{10}$, and $10^{11}$ \Msun, respectively;
see \Fig\ref{fig:density_vs_z}). Notice that objects in the
\cite{Damjanov+14_compacts} sample are not classified according to
either morphology or galaxy age, nor do they have an accurate
estimate of the intrinsic \Re.  Hence, their selection may be
missing compact systems that are actually spatially resolved in
BOSS-DR10. On the other hand, since \cite{Damjanov+14_compacts}
also include compacts with $\Re > 1.5 \, \rm kpc$ and are selected
with respect to \Mdyn, rather than \mst , it is not surprising,
perhaps, that those abundance estimates are larger than ours. In
fact, if we consider the abundance estimate of massive BOSS
targets with $\Re<1.5$ kpc and $\Mdyn > 8 \times 10^{10}\, \rm
\Msun$ from \cite{Damjanov+14_compacts} (see purple triangle in
the left panel of \Fig\ref{fig:density_vs_z}), the number density
of compacts in the lowest redshift bin ($z \sim 0.25$) is $(1.7
\pm 0.5) \times 10^{-6}\, \rm Mpc^{-3}$, in much better agreement
with our density estimate of $(1.6 \pm 0.4) \times 10^{-6}\, \rm
Mpc^{-3}$ for the same redshift bin.  On the other hand, selecting
galaxies with $\Re < 2.5$ kpc in KiDS, would lead to abundances
$\sim 0.8$ dex larger than for $\Re < 1.5$ kpc, still in good
agreement with estimates for stellar-like objects from
\cite{Damjanov+14_compacts}. The yellow region in the left panel
of \Fig\ref{fig:density_vs_z} plots number densities for galaxies
in the COSMOS survey \citep{Damjanov+15_compacts}\footnote{ These
data are kindly computed for us by I. Damjanov (private
comunication) by applying the same size and mass selection
criteria as in the present work.}. Remarkably, no evolution with
redshift is found, for both KiDS and COSMOS samples, although,
surprisingly, abundances for COSMOS (on average $\sim 10^{-5}\,
\rm Mpc^{-3}$) are about one order of magnitude larger than our
KiDS estimates.  Since \cite{Damjanov+14_compacts} claim to find
consistent density estimates between COSMOS and BOSS (the latter
having an area 4000 times larger than COSMOS), cosmic variance
seems not to be responsible for the above discrepancy. However, we
cannot exclude that galaxy environment, which might be the actual
driver of the number density of compact relics at $z \sim 0$ (see,
e.g. \citealt{Trujillo+14}; \citealt{Poggianti+13_low_z};
\citealt{Valentinuzzi+10_WINGS}), may be different for galaxies in
the KiDS-DR2 and COSMOS areas -- an issue that will be addressed
in forthcoming extensions of the present work.

The  results  for photometrically  old  \MSCGs\ (right  panel of
\Fig\ref{fig:density_vs_z})   are  first  compared with estimates
for $z_{f}>2$  compact  galaxies from \cite{Damjanov+14_compacts},
who selected  samples with two cuts in \Mdyn\ ($> 6 \times
10^{10}$ and $8 \times 10^{10}$, respectively; see
\Fig\ref{fig:density_vs_z}). As for the  left panel, there  are
differences in the  selection criteria that make a direct
comparison  not trivial. However, the  trends for both BOSS and
KiDS  samples look qualitatively  similar, with  a very mild
evolution  with redshift in  the range  where our KiDS sample is
complete. Moreover, as noticed above, for BOSS compacts with $\Re
< 1.5  \rm kpc$, the \cite{Damjanov+14_compacts}  abundance
estimates are fairly consistent  with ours, within the
uncertainties.

Finally, we compare  \MSCGs\ number densities  with predictions
from semi-analytical models~\footnote{We caution the reader that
stellar masses and sizes are measured in a different way between
simulations and observations, hampering a straightforward
comparison of the two.} (SAMs). \cite{Quilis_Trujillo13} have
determined the evolution of the abundance of compact galaxies from
SAMs based on Millennium N-body simulations (\citealt{Guo+11_sims,
Guo+13_sims}), where relic compacts are defined as galaxies which
have barely increased their stellar mass between $z\sim 2$ and
$z\sim 0$.  Operatively, they selected from the merger tree those
objects that have increased their mass since $z=2$ by less than
10\% and 30\% , respectively, i.e.  galaxies whose  mass at $z
\sim 2$ is larger than 90\% and 70\% of the mass limit applied to
select compacts.  Our results are consistent in the lowest
redshift bin with \cite{Guo+13_sims} for simulated galaxies which
have increased their mass at most by $10\%$, while we estimate
lower densities than in simulations at higher redshift. However,
similarly to what discussed for the comparison to BOSS estimates,
theoretical predictions should be actually considered as upper
limits, as \cite{Quilis_Trujillo13} did not apply any precise
selection in size, since the resolution in the simulations does
not allow reliable estimates of galaxy effective radii to be
obtained. We also compare our findings to results from the
hydrodynamical Illustris simulation of \cite{Wellons+15_lower_z}
(see also \citealt{Wellons+15_z2}). \cite{Wellons+15_lower_z}
select 35 massive and compact galaxies at $z=2$ and follow their
evolution to $z=0$. Only 1 out of these 35 systems evolves into a
galaxy that satisfies our mass and size criteria at $z \leq 0.5$.
This corresponds to a number density of $8.28 \times 10^{-7}\, \rm
Mpc^{-3}$, which is consistent with the abundances of compact
galaxies that have accreted less than 10\% of their final mass
from the Millenium simulations. As a further caveat here, we point
out that in our selection we adopt the same mass cutoff value at
all redshifts, while simulations perform the mass selection at
$z=0$, implying that at high redshifts, simulated galaxies with
masses smaller than the $z=0$ cutoff value are actually included
in the analysis. This is another reason why one may expect that
predicted number densities for compacts at high redshift are
actually higher than the observed ones.

At redshifts $z \lsim 0.2$, in the left panel of
\Fig\ref{fig:density_vs_z}, we see a lack of candidates.  This
seems to contrast the results of \citet{Trujillo+09_superdense}
who found 29 secure \MSCGs\ at $z<0.2$ fulfilling  our same
criteria, all of them having young ages $\lsim 4\, \rm Gyr$ (see
also \citealt{Ferre-Mateu+12}). However, one should notice that
out of the 29 \MSCGs\ of \citet{Trujillo+09_superdense}, only one
is at redshift $<0.1$, still pointing to the extreme paucity of
such systems in the nearby Universe, consistent with our result.
Similarly, \cite{Taylor+10_compacts}  found one possible old
\MSCG\ at low redshift, using a more relaxed criterion for the
size, than the one we adopt here.

\cite{Saulder+15_compacts} have found a sample of 76 compact
galaxies from SDSS at $0.05 < z < 0.2$, which resemble quiescent
galaxies at high-z, i.e. systems with small effective radii and
large velocity dispersions. If we consider their compacts with
$\Re<1.5$ kpc and $\mst > 8 \times 10^{10}\, \rm \Msun$, 1 galaxy
at $z<0.1$ and 6 galaxies at $z>0.1$ are left when the effective
radius from a de Vaucouleurs profile is used. Instead, these
numbers change to 1 galaxy at $z<0.1$ and only 1 at $z>0.1$ if a
S\'ersic profile is fitted. These numbers correspond to abundances
of $2.4\times 10^{-8}\, \rm Mpc^{-3}$ in the redshift range $0.05
< z < 0.1$, and $2\times 10^{-8}\, \rm Mpc^{-3}$ and $3.3\times
10^{-9}\, \rm Mpc^{-3}$ at $0.1 < z < 0.2$, if de Vaucouleurs or
S\'ersic profile are fitted, respectively. As mentioned in
\Sec\ref{sec:intro}, these findings seem to trouble the current
hierarchical paradigm of galaxy formation, where some relic
systems at $z \sim 0$ are actually expected to be found. In
contrast, \cite{Poggianti+13_low_z} have found 4 galaxies
fulfilling our same criteria (corresponding to $1.4\%$ of their
sample galaxies with masses larger than $8 \times 10^{10}\, \rm
\Msun$), and all of these galaxies are old, with mass-weighted
ages older than $8\, \rm Gyr$. These numbers translate into a very
large abundance of $\sim 10^{-5}\, \rm Mpc^{-3}$. Recently,
\cite{Trujillo+14} added a new brick to the story, finding one
relic compact in the Perseus cluster, i.e. NGC 1277, reconciling
the observations at $z \lsim 0.2$ with predictions from
simulations. All these results point to an overabundance of
\MSCGs\ in dense cluster regions, that are actually
under-represented over the area currently mapped by KiDS.

As discussed in \Sec\ref{subsec:wrong_z},  the sample of \MSCGs\
would be  reduced significantly  in  size  by systematics in the
redshift determination. \Fig\ref{fig:density_vs_z} ``corrects''
the abundance estimates for the  possible systematics in redshift.
The corrected abundances (see gray boxes and lines in
\Fig\ref{fig:density_vs_z}) should be seen as our current lower
limit on number densities of SMCGs in KiDS. This issue  will be
addressed by on-going spectroscopic follow-up of \MSCGs, and
discussed in details in a future paper.

\section{Conclusions}\label{sec:conclusions}

Thanks to the large area covered, high image quality, excellent
spatial resolution and seeing, the Kilo Degree Survey (KiDS) provides
a unique dataset to study the properties of super-compact massive
galaxies (\MSCGs) --  a family of systems which plays a key role into our
understanding of galaxy formation and evolution.

In this paper, we present a sample of candidate \MSCGs\ , based on
156~sq.~deg. of KiDS, in four optical bands (\u, \g, \rband\ and
\i). We start from a sample of $\sim 0.4$ million galaxies with
high-\SN, measured photometry and structural parameters in all the
four bands. For a subsample of these galaxies, we have used the
KiDS photometry to estimate: 1) photometric redshifts based on
machine learning techniques (\citealt{Brescia+14};
\citealt{Cavuoti+15_KIDS_I}); 2) structural parameters using the
software 2DPHOT (\citealt{LaBarbera_08_2DPHOT}); 3) stellar
masses, fitting colours with SPS model predictions. The resulting
sample is $>90\%$ complete down to an \rband-band magnitude $\sim
21$, and down to a stellar mass of $3-5 \times 10^{10} \, \rm
\Msun$, up to a redshift $z \sim 0.5$. We select the most massive
($\mst
> 8\times 10^{10} \, \rm$) and most compact ($\Re < 1.5 \, \rm
kpc$) galaxies with (photometric) redshift $z \lsim 0.7$. We remove
star contaminants by performing a visual inspection of the final sample of
candidates and then, for galaxies with available near-IR
photometry from VIKING-DR1, we combine optical+NIR photometry to reduce
the fraction of contaminants.

The final sample consists of 92 compact candidates, with a number
density of $\sim 0.9$ compact galaxies per square degree, at
$z<0.7$. Nine candidates have spectroscopic information from SDSS
and GAMA surveys, that is used to assess the systematics in the
redshift determination of our sample.  On average, compact
galaxies have negative colour gradients which are similar to the
ones for normal passively evolving galaxies (e.g.,
\citealt{Tamura+00}; \citealt{Tamura_Ohta00};
\citealt{Tortora+10CG}; \citealt{Tortora+13_CG_SIM};
\citealt{LaBarbera+12_SPIDERVII_CG}). However, the variety of
gradients, to be confirmed with a spectroscopic follow-up of the
present sample, seems to suggest that compact galaxies formed
under a wide range of initial conditions (\citealt{Gargiulo+12}).
We also discuss the evolution with redshift of the number density
of compact systems, and in particular that for the oldest
galaxies, which are possibly remnants of the high-redshift ($z_{f}
\gsim 2$) compact population detected by many studies. Remarkably,
we do not find any \MSCGs\ candidate at $z \lsim 0.2$.  This
finding, which is consistent with the recent studies, might be
related to the effect of galaxy environment on the abundance of
compact systems (see Sec.~1). Although observational studies at
intermediate- and high-z do not point to a clear picture of how
environment affects galaxy sizes (see
\citealt{Damjanov+15_env_compacts} and references therein), recent
cosmological simulations predict a larger fraction of massive
compact systems in high- than in low-density regions
(\citealt{Stringer+15_compacts}).  This prediction is supported by
the fact that NGC\,1277 -- the only well-characterized compact,
massive, ETG at $z \sim 0$ -- resides in the core region of a
nearby rich cluster of galaxies (\citealt{Trujillo+14}; see also
\citealt{Valentinuzzi+10_WINGS}; \citealt{Poggianti+13_low_z}).
Thus, the absence of compact galaxies at $z < 0.2$ could be
related to the smaller fraction of dense structures in the area
currently mapped by KiDS.

We plan to have 10 times more compact candidates at the end of the
KiDS survey, when all the 1500~sq.~deg. will have been observed.
Only nine of our candidates have a spectroscopic coverage at the
moment. Thus, a spectroscopic follow-up is actually necessary, to
fully validate and characterize our sample, and determine accurate
number densities to be compared with theoretical expectations.

%%%%%%%%%%%%%%%%%%%%%%%%%%%%%%%%%%%%%%%%%%%%%%%%%%%%%%%%%%%%%%%%%%%%%%%

\section*{Acknowledgments}

Based on data products from observations made with ESO Telescopes
at the La Silla Paranal Observatory under programme IDs
177.A-3016, 177.A-3017 and 177.A-3018, and on data products
produced by Target/OmegaCEN, INAF-OACN, INAF-OAPD and the KiDS
production team, on behalf of the KiDS consortium. OmegaCEN and
the KiDS production team acknowledge support by NOVA and NWO-M
grants. Members of INAF-OAPD and INAF-OACN also acknowledge the
support from the Department of Physics \& Astronomy of the
University of Padova, and of the Department of Physics of Univ.
Federico II (Naples). CT has received funding from the European
Union Seventh Framework Programme (FP7/2007-2013) under grant
agreement n. 267251 ``Astronomy Fellowships in Italy'' (AstroFIt).
This work was partially funded by the MIUR PRIN Cosmology with
Euclid. MB acknowledges the PRIN-INAF 2014 {\it Glittering
kaleidoscopes in the sky: the multifaceted nature and role of
Galaxy Clusters}. We thank the referee for his/her comments, which
have helped to improve the paper. We thank I. Damjanov for having
provided us with the abundances of compacts for the COSMOS field,
applying the same size and mass selections as in the present work.
We thank M. Cacciato, R. C. E. Van den Bosch, M. Viola and I.
Trujillo for helpful comments, B. Poggianti for comments about
their findings at low redshift, V. Quilis who provided us with
simulation predictions. We also thank S. Wellons and A. Pillepich
for comments about their Illustris simulations.

%%%%%%%%%%%%%%%%%%%%%%%%%%%%%%%%%%%%%%%%%%%%%%%%%%%%%%%%%%%%%%%%%%%%%%%

\bibliographystyle{mn2e}   % (uses fill "plain.bst")

%\bibliography{C:/Users/crescenzo/Documents/latex/Bibtex/myrefs,C:/Users/crescenzo/Documents/latex/Bibtex/myrefs_KiDS_add}       % expects file "myrefs.bib"

%\bibliography{D:/Documenti/Lavoro/latex/Bibtex/myrefs}       % expects file "myrefs.bib"

\end{document}